\documentclass{PoS}
\usepackage{graphicx}                
\usepackage{bm}                      
\usepackage{mathptmx}                
\usepackage{xcolor}                  

\def\bea{\begin{eqnarray}} \def\eea{\end{eqnarray}}
\def\beq{\begin{equation}} \def\eeq{\end{equation}}
\def\bal#1\eal{\begin{align}#1\end{align}}
\def\bse#1\ese{\begin{subequations}#1\end{subequations}}

\def\text{\mathrm}
\def\be{\beta}

\def\la{\Lambda}
\def\tl{\tilde\Lambda}
\def\ms{M_\odot}
\def\mmax{M_\text{max}}

\title{Constraints from the GW170817 merger event
on the nuclear matter equation of state}
\ShortTitle{Constraints from the GW170817 merger event}

\author{\speaker{G. F. Burgio}\\
INFN Sezione di Catania, Via Santa Sofia 64, 95123 Catania, Italy\\
E-mail: \email{fiorella.burgio@ct.infn.it}}

\author{A. Figura\\
Dipartimento di Fisica, Universit\'a di Catania, and INFN Sezione di Catania,
Via Santa Sofia 64, 95123 Catania, Italy\\
E-mail: \email{figura.antonio@gmail.com}}

\author{H.-J. Schulze\\
INFN Sezione di Catania, Via Santa Sofia 64, 95123 Catania, Italy\\
E-mail: \email{schulze@ct.infn.it}}

\author{J.-B. Wei\\
Dipartimento di Fisica, Universit\'a di Catania, and INFN Sezione di Catania,
Via Santa Sofia 64, 95123 Catania, Italy\\
E-mail: \email{jinbiao.wei@ct.infn.it}}

\abstract{
The detection of the GW170817 neutron star merger event has incited
an intense research activity towards the understanding of the
nuclear matter equation of state.
In this paper we compare in particular the pressure-density relation
obtained from heavy-ion collisions with the analysis of the  NS merger event.
Moreover, we present recent calculations of neutron star's moment of inertia
and tidal deformability using various microscopic equations of state
for nuclear and hybrid star configurations,
and confirm several universal relations.
We also discuss the recent constraints for the NS radii determined by GW170817,
and find compatible radii between 12 and 13 kilometers,
thus identifying the suitable equations of state.
}

\FullConference{
XIII Quark Confinement and the Hadron Spectrum - Confinement2018\\
31 July - 6 August 2018\\
Maynooth University, Ireland}

\begin{document}

\section{Introduction}

Neutron star (NS) observations allow us to explore the equation of state (EOS)
of nuclear matter \cite{eos} at densities well beyond the ones
available in terrestrial laboratories.
Currently the masses of several NSs are known with good precision \cite{mass},
but the information on their radii is not very accurate
\cite{ozel16,gui2013,lat2014}.
Present and future observations of the NS features,
such as the mass--radius relation \cite{lattimer2016}
and the mass--moment-of-inertia relation \cite{schutz},
can help to infer the NS EOS within a certain observational uncertainty
in the upcoming years.

The recent detection by the Advanced LIGO and Virgo collaborations
of gravitational waves emitted during the GW170817 merger event \cite{merger}
has provided important new insights on the structural properties
of these objects,
most prominently their masses and radii,
by means of the measurement of the tidal deformability \cite{hartle,flan},
and to deduce upper \cite{merger} and lower \cite{radice} limits on it.
In this paper we examine these quantities and their relations
with other observables \cite{yagi13,yagi17},
using several EOSs, both microscopic and phenomenological ones \cite{2018Chap6}.

The paper is organized as follows.
In Sec.~2 we give a brief overview of the hadronic and hybrid EOSs we
are using.
In Sec.~3 we discuss a comparison of the EOSs with the one
from the GW merger event and heavy-ion collision data,
along with the numerical results obtained for the NS mass-radius relation,
and the well-known universal relations among the tidal deformability,
the moment of inertia, and the quadrupole moment.
In Sec.~4 we draw our conclusions.

\section{Equations of state}

The various EOSs used in this paper are mainly microscopic EOSs based on the
Brueckner-Hartree-Fock (BHF) many-body theory with realistic
two-body and three-body nucleonic forces \cite{Jeu1976,baldo1999}.
In particular we examine several EOSs \cite{li08}
based on different nucleon-nucleon potentials,
the Argonne $V_{18}$ (V18, UIX) \cite{v18},
the Bonn B (BOB) \cite{bonn1,bonn2},
and the Nijmegen 93 (N93) \cite{nij1,nij2},
and compatible three-nucleon forces \cite{glmm,uix3,tbfnij} as input.
For completeness, we also compare with the often-used results of the
Dirac-BHF method (DBHF) \cite{dbhf3},
which employs the Bonn~A potential,
and the APR EOS based on the variational method \cite{apr1998}
and the $V_{18}$ potential.
Two phenomenological relativistic-mean-field EOSs are also used for comparison:
LS220 \cite{ls} and SFHo \cite{sfh}.

As far as the hybrid-star EOS is concerned,
since the EOS for quark matter (QM) is poorly known at zero temperature
and at the high baryonic density appropriate for NSs,
one can presently only resort to more or less phenomenological models
for describing QM,
such as the MIT bag model \cite{Chod} or the Nambu--Jona-Lasinio model
\cite{Buballa2005,klahn2015}.
We adopt an alternative approach based on the Dyson-Schwinger equations,
which provide a nonperturbative continuum field approach to QCD
that can simultaneously address both confinement and dynamical
chiral symmetry breaking \cite{Roberts1994,Alkofer2000wg}.
In Refs.~\cite{chen11,chen17a}
we developed a Dyson-Schwinger model (DSM) for deconfined QM,
which was combined with the BHF approach for the hadronic phase
in order to model NSs.
The quark phase in the DSM is modeled with an interaction parameter
$\alpha=1,2$,
and the corresponding models are labelled as DS1 and DS2.
The construction of the Gibbs phase transition is then performed for each of the
nucleonic models.
This yields several EOSs which differ essentially by their onset density
of the QM phase and the associated NS maximum mass.
We stress that those combinations,
labeled by BOB+DS1, BOB+DS2, V18+DS1, and N93+DS1,
give a value of the static maximum mass larger than two solar masses,
thus compatible with observations \cite{demorest2010,fonseca2016,heavy2}.

Properties of the various NS configurations constructed with the considered EOSs
are listed in Table~\ref{t:NSlist},
namely, the value of the maximum mass, the corresponding radius,
the radius of the $1.4\,\ms$ configuration
and its tidal deformability $\la_{1.4}$,
which will be extensively discussed in Sect.~3.

We mention that the BHF theory was also extended in order to include hyperons,
which might appear in the core of a NS,
but the corresponding EOSs turn out to be very soft,
with too low NS maximum masses,
$M<1.7\,\ms$ ($\ms\approx2\times10^{33}$g) \cite{mmy1,mmyy,mmy2},
well below the current observational limit.
Nevertheless, such EOSs could be realized in the so-called two-families scenario
in which the heaviest stars are interpreted as quark stars,
whereas the lighter and smaller stars are hadronic stars
\cite{drago1,drago2,drago4,pascha}.
Those are labelled as V18(N+Y) and BOB(N+Y) in Table~\ref{t:NSlist}.

\def\myc#1{\multicolumn{1}{c}{$#1$}}
\begin{table}
\renewcommand{\arraystretch}{0.9}
\begin{center}
\caption{
\label{t:NSlist}
Properties of NSs listed according to the considered EOSs.
See text for details.}
\medskip
\begin{tabular}{@{}lcccr}
\hline
  EOS & \myc{\mmax[\ms]} & $R_{\mmax}$ [km] & $R_{1.4}$ [km] &
  $\phantom{-.}\la_{1.4}$ \\
\hline\\[-4mm]
  BOB      & 2.51 & 11.32 & 12.85 & 584 \\
  BOB+DS1  & 2.30 & 12.13 & 12.85 & 584 \\
  BOB+DS2  & 2.02 & 11.95 & 12.85 & 584 \\
  V18      & 2.34 & 10.63 & 12.33 & 419 \\
  V18+DS1  & 2.16 & 11.34 & 12.33 & 419 \\
  N93      & 2.13 & 10.49 & 12.68 & 474 \\
  N93+DS1  & 2.00 & 11.17 & 12.68 & 474 \\
  UIX      & 2.04 & 10.02 & 12.03 & 340 \\
  APR      & 2.20 &  9.92 & 11.59 & 274 \\
  DBHF     & 2.31 & 11.29 & 13.10 & 681 \\
  SFHO     & 2.06 & 10.31 & 11.93 & 334 \\
  LS220    & 2.04 & 10.67 & 12.94 & 542 \\
  V18(N+Y) & 1.65 &  9.00 & 11.92 & 302   \\
  BOB(N+Y) & 1.37 & 11.07 &  $-$  & $-$  \\
\hline
\end{tabular}
\end{center}
\end{table}

\begin{figure}[t]
\vspace{0mm}
\centerline{\includegraphics[scale=0.6]{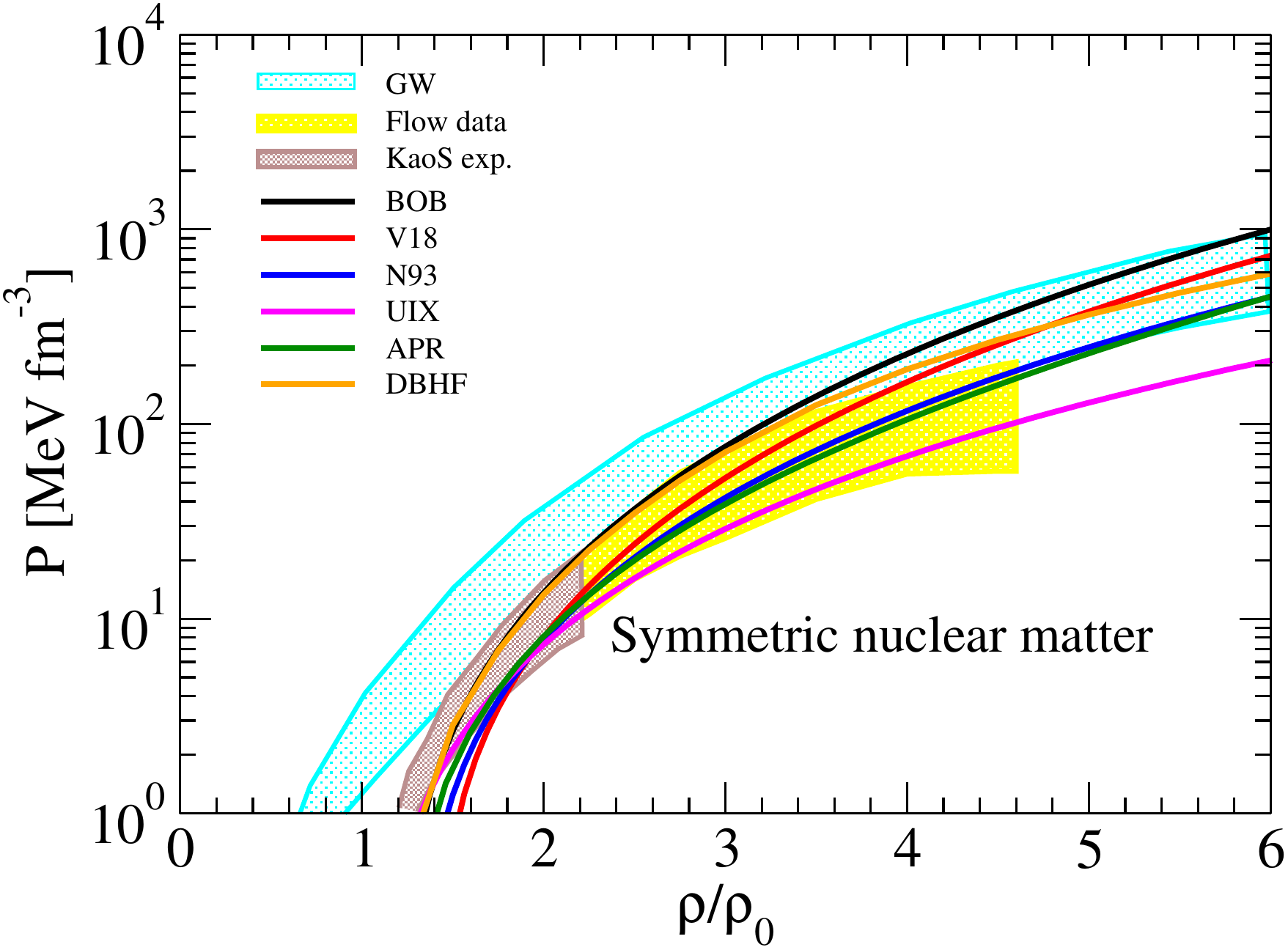}}
\vspace{-2mm}
\caption{
Pressure vs.~density of symmetric nuclear matter for
different EOSs in comparison with the phenomenological constraints
from heavy ion collisions (brown and yellow shaded areas),
and from GW170817 (blue shaded area).
$\rho_0=0.16$ fm$^{-3}$ is the saturation density.
}
\label{f:flow}
\end{figure}

\section{Results and discussion}

\subsection{Constraints on the EOS from heavy-ion collisions
and gravitational waves}

In the last two decades intensive studies of heavy ion reactions at energies
ranging from few tens to several hundreds MeV per nucleon have been performed in
different nuclear physics laboratories.  The main goal has been the extraction
of the gross properties of the nuclear EOS from the data.
In fact, in heavy ion collisions at large enough energy,
nuclear matter is compressed and,
at the same time, the two collision partners produce flows of matter, i.e.,
a large abundance of nucleons and fragments of different sizes are emitted.
One observable that is often analyzed is the transverse flow,
which depends sensitively on the pressure developed in the fireball, i.e.,
the center of the interaction zone at the moment of maximum compression
during the collision.

On the other hand,
also subthreshold $K^+$ production in heavy ion reactions has been demonstrated
to probe the fireball density reached during the collision,
being this the ideal situation for exploring the EOS and its incompressibility.
In Ref.~\cite{daniel2002} the flow and kaon production analysis
was summarized by plotting the region in the pressure vs.~density plane
where any reasonable EOS should pass through,
using all the simulations with different EOSs compatible with different data.
That analysis is displayed in  Fig.~\ref{f:flow} as a hatched (yellow) box
for the flow data by the FOPI \cite{fopi} collaboration,
and as a (brown) box for the subthreshold kaon production by
the KaoS \cite{kaoS} collaboration.
Those results point in the direction of a soft EOS,
i.e., values of the compressibility $K$ in the range
$180 \leq K \leq 250$ MeV at density close to saturation.
Those values are compatible with the ones extracted from the data on
monopole oscillations \cite{colo2004}.

The EOS governs not only the dynamics of heavy ion collisions,
but also that of neutron star mergers.
For instance, the final fate of the merger, i.e.,
prompt or delayed collapse to a black hole or a single neutron star,
does depend on the EOS,
as well as the amount of ejected matter which undergoes nucleosynthesis
of heavy elements.
During the inspiral phase,
the influence of the EOS is evident on the tidal polarizability,
$\Lambda = \frac{2}{3} k_2 \beta^{-5}$,
where $k_2$ is the Love number and
$\beta=M/R$ is the compactness.

An upper limit of $\Lambda<800$ was initially given in the
first GW170817 analysis for a $1.4\,\ms$ neutron star \cite{merger},
but later on the analysis was improved by assuming that both neutron stars
have the same EOS,
thus giving different limits of
$\Lambda=190^{+390}_{-120}$ and
$\rm R=11.9^{+1.4}_{-1.4}\;\text{km}$ \cite{Abbott:2018exr}.
In this latter analysis,
the values of the pressure as a function of density were extracted,
and those are displayed as the blue hatched area in Fig.~\ref{f:flow}.
We notice that almost all EOSs, except UIX,
turn out to be compatible with the GW170817 data at density $\rho>2\rho_0$,
whereas the nuclear collision data look more restrictive.
In fact, among the considered EOSs,
UIX, APR, and N93 are well compatible with the data
extracted from heavy ion collisions over the whole density range,
whereas BOB, DBHF, and V18 are only marginally compatible at large density,
and those are characterized by a large stiffness.
A further analysis of heavy ion collision data,
compared to the ones from GW observations,
can be found in Ref.~\cite{2018arXiv180706571T}.

\begin{figure*}[htbp]
\parbox{0.5\hsize}{
\includegraphics[width=\hsize]{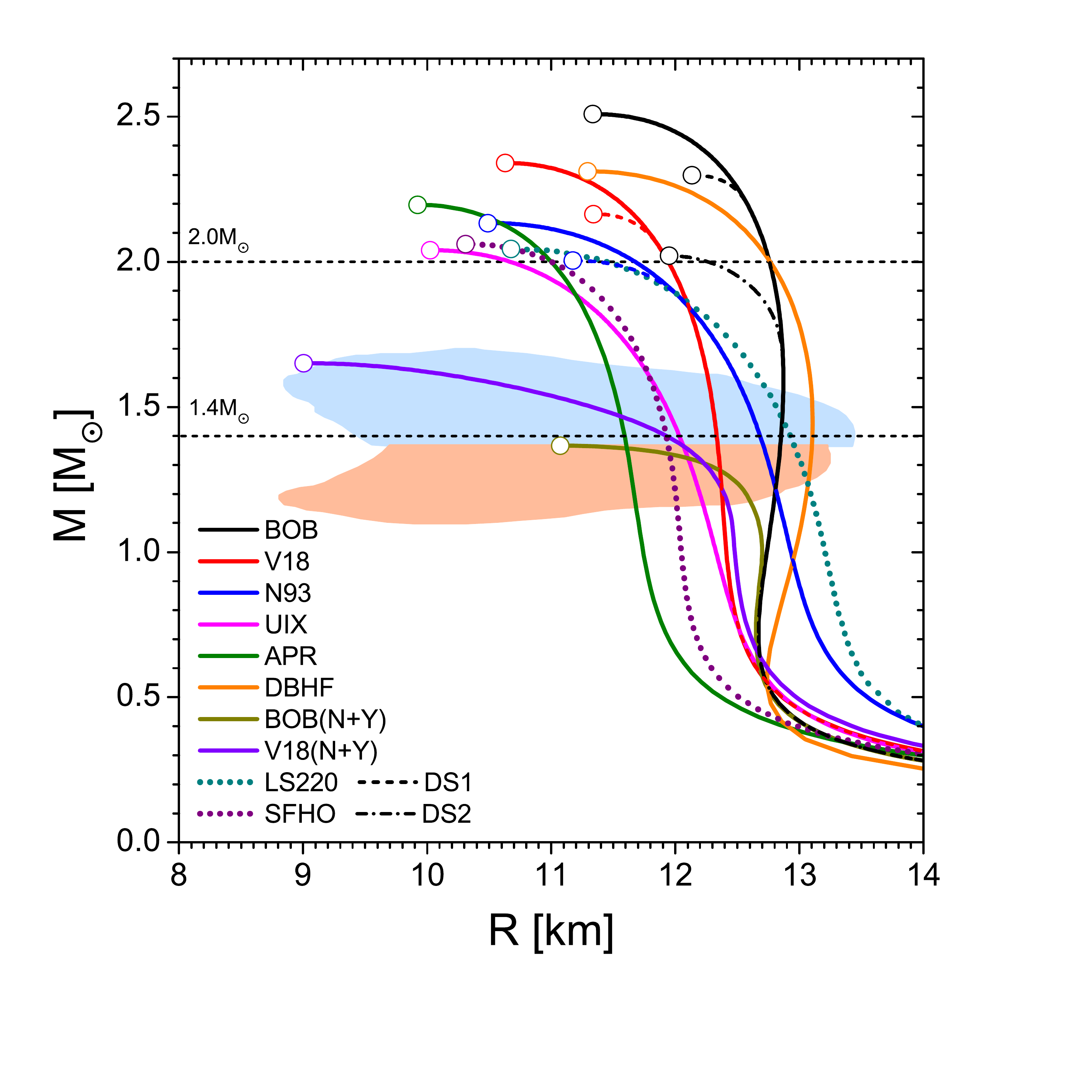}}
\hskip-20mm
\parbox{0.5\hsize}{
\includegraphics[width=0.88\hsize,angle=90]{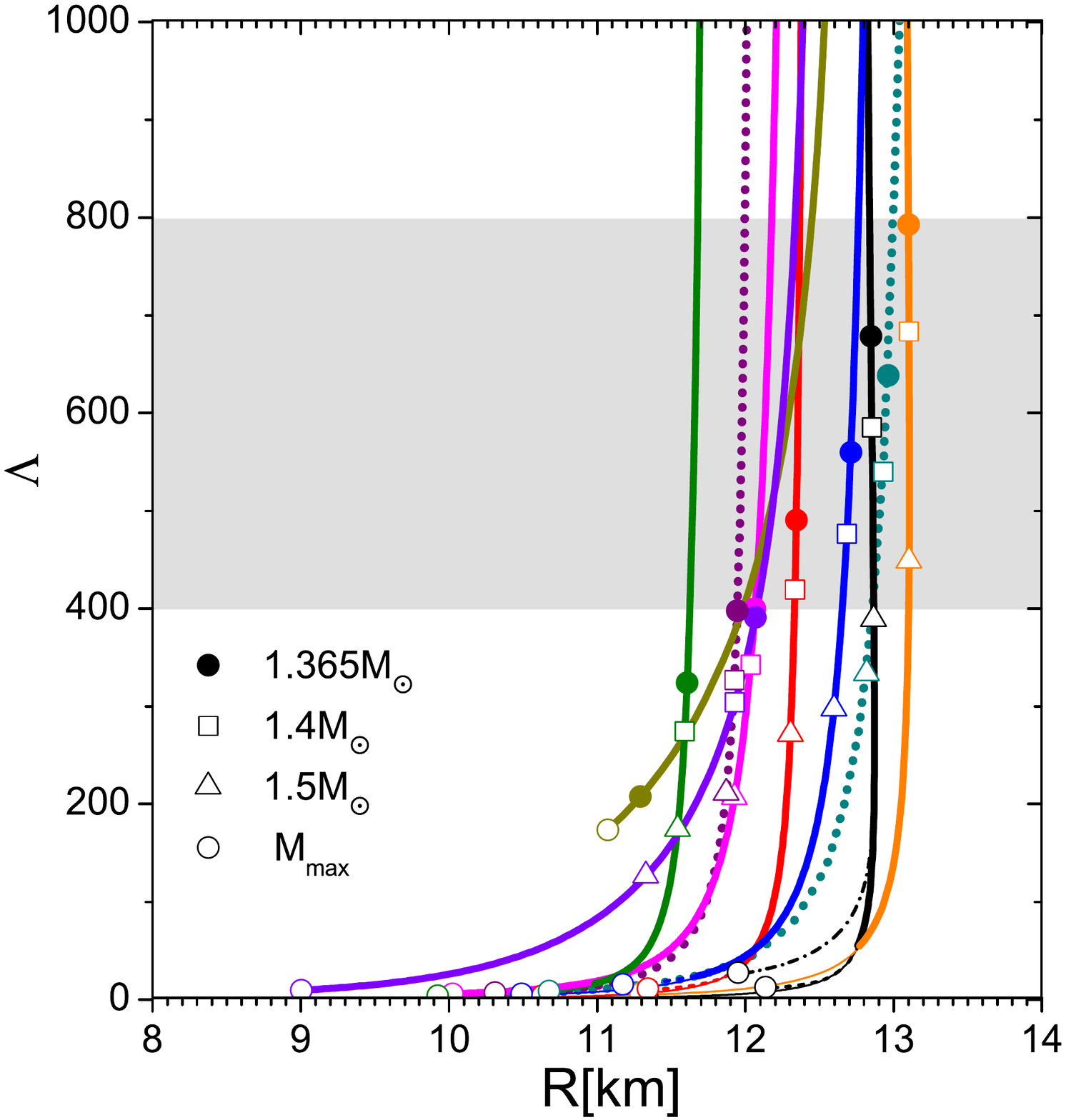}\\[3ex]
}
\vspace{-7mm}
\caption{
(Left panel) Mass-radius relations for different EOSs.
Solid (dotted) curves are plotted for microscopic (phenomenological) EOSs.
Dashed and dot-dashed curves display hybrid stars in the DSM approach
with DS1 and DS2, respectively.
Open circles indicate the values of the maximum mass.
The shaded areas show limits derived in \cite{Abbott:2018exr}.
(Right panel) Correlations between $M$, $R$, and $\la$
for a single NS with different EOSs.
The shaded area is constrained by the interpretation of the GW170817 event
as a symmetric NS merger.
}
\label{f:mr}
\end{figure*}

\subsection{Mass-radius relations and deformability}

A very important constraint to be fulfilled is the value of the maximum mass
for the different EOSs,
which has to be compatible with the observational data.
In Fig.~\ref{f:mr} (left panel) we display the mass-radius relations
obtained with microscopic and phenomenological EOSs,
shown respectively as solid and dotted lines.
Moreover we consider the four EOSs for hybrid stars
with $\mmax>2\ms$ in Table~\ref{t:NSlist},
which are displayed as dashed and dot-dashed lines, respectively.
We observe that most models give values of the maximum mass
larger than $2\,\ms$,
and therefore are compatible with current observational data
\cite{demorest2010,fonseca2016,heavy2},
except the two hyperonic EOSs.
Some recent analyses of the GW170817 event
indicate an upper limit of the maximum mass of about $2.2\,\ms$
\cite{shiba17,marga17,rezz18},
with which several of the microscopic EOSs would be compatible.

Regarding the radii,
according to Fig.~\ref{f:mr} the predicted radii for a $M=1.4\,\ms$ NS
span a range
$(11.6 \lesssim R_{1.4} \lesssim 13.1)\,$km,
fully compatible with the recent analysis in \cite{Abbott:2018exr}
(shaded areas),
and also in agreement with Ref.~\cite{annala17},
where it has been shown that the tidal deformability limit
of a $1.4\,\ms$ NS,
$\la_{1.4} < 800$, as found in GW170817,
implies a radius $R_{1.4}<13.6\;$km.
Quite similar upper limits have been obtained in \cite{most18,lim18,raithel18}.

The interpretation of the GW170817 event
also allowed to establish {\em lower} limits on the NS radius.
The condition of (meta)stability of the produced hypermassive star after merger
allowed to exclude very soft EOSs \cite{shiba17}
and to set thus a lower limit on the radius,
$R_{1.6} > 10.7\,$km \cite{baus17},
confirmed by similar recent analyses \cite{most18,lim18} in which
$R_{1.4}>(11.5-12)\,$km.
Simulations with several different EOSs set also a lower limit
on the effective deformability, $\tl>400$ \cite{radice},
related to the black hole formation time
and the accretion disk mass of material left out of the black hole.
The latter was constrained from optical/infrared observations
of the remnant AT2017gfo \cite{knova1,knova2,knova3,knova4,knova5}.
Small values of $\tl$ and therefore small values of $R$ imply
very fast black hole formation and little material left in the disk,
which is incompatible with observation.

On the other hand, radii smaller than these lower limits were deduced
from observations of thermal emission from accreting NSs in quiescent LMXBs,
which seem to suggest for stars of mass about $(1.4-1.5)\,\ms$
a radius in the range $(9.9-11.2)\,$km \cite{ozel16}.
Those results have been criticized in
\cite{lattimer2001,steiner2010,steiner2013}:
in particular the estimates of the radii are affected by the uncertainties
on the composition of the atmosphere. At the moment,
no firm conclusions can yet be reached and we need to wait
for new data such as the ones collected by the NICER mission,
in order to obtain independent and precise information on NS radii.
We remark that this clash between large radii from GW170817
and small radii from quiescent LMXBs (if confirmed)
could be resolved in the two-families or twin-star scenarios,
in which small and big stars of the same mass
could coexist as hadronic and QM stars
\cite{drago1,drago2,drago4,pascha,drago3}.

Fig.~\ref{f:mr} (right panel) shows the tidal deformability $\la$
as a function of $M$ and $R$ for the various EOSs considered in this paper.
We remind the reader that the tidal deformability $\la$ can be computed
by numerically solving for the interior and exterior gravitational field of a NS
in a slow-rotation \cite{hartle} and small-tidal-deformation approximation
\cite{hinder2008,hinder2009,hinder2010}.
The information on the mass $M$ is encoded in the various symbols
along the curves.
The grey band represents the observational limits
derived in Refs.~\cite{merger,radice}
mentioned before, i.e., $400<\la<800$.
The full dots represent the masses  $M=1.365\,\ms$
of each NS for a symmetric binary system in GW170817,
whereas the empty symbols represent respectively the canonical mass
$M=1.4\,\ms$ (squares),
and the maximum mass $M=\mmax$ (open circles).
The empty diamond corresponds to $M=1.5\,\ms$ and represents the constraint
derived in Ref.~\cite{annala17}.
One notes that the conditions $M=1.365\,\ms$ and $400<\la<800$
imply $12\,$km$\,\lesssim R \lesssim 13\,$km,
with the compatible EOSs V18(N+Y), UIX, V18, N93, BOB, DBHF
in order of increasing radius.
Also the phenomenological EOS labelled LS220 fullfills the constraint,
as well as trivially the hybrid stars constructed with DS1 and DS2 EOSs,
which at $M=1.365\,\ms$ are still purely nucleonic.
On the other hand, APR, BOB(N+Y), and SFHO (marginally)
do not fulfill the $\la>400$ constraint.

\subsection{Universal relations}
\label{subs:form}

Universal (EOS-independent) relations between the NS moment of inertia $I$,
the NS Love number $k_2$, and the (spin-induced) NS quadrupole moment $Q$
(I-Love-Q relations) have been widely discussed in the last few years
\cite{yagi13,yagi17},
and they turn out to have various useful applications in astrophysics,
because the measurement of any member of the I-Love-Q trio automatically
predicts the remaining two quantities without having to know the EOS.
The tidal Love number, for example,
has been measured by Advanced LIGO and Virgo collaboration \cite{merger},
and by combining this measurement with the I-Love-Q relations,
one can obtain the moment of inertia and the quadrupole moment of NSs
in a binary system,
which would also be difficult to measure from GW observations.
Provided the validity of a universality relation,
the moment of inertia $I$ of a NS can be expressed as a function
of the NS mass and radius,
and therefore the radius could be determined if the mass and the
moment of inertia of the NS is known \cite{schutz,wor}.
Below we discuss some of those relations.
For details the reader is referred to Ref.~\cite{2018arXiv180904315W}.

\begin{figure}[t]
\vspace{-9mm}\hspace{-2mm}
\includegraphics[scale=0.284]{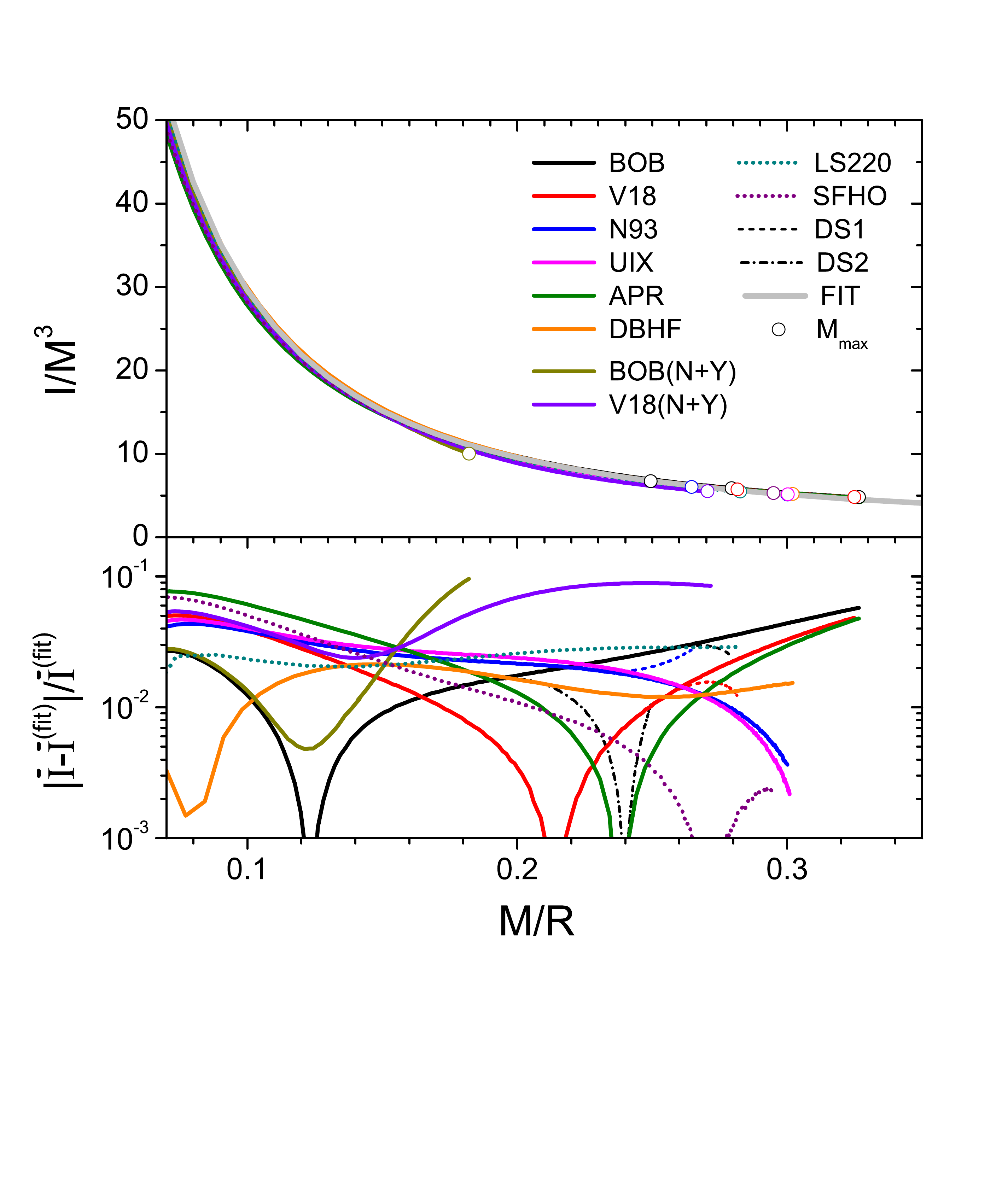}
\hskip-9mm
\includegraphics[scale=0.293]{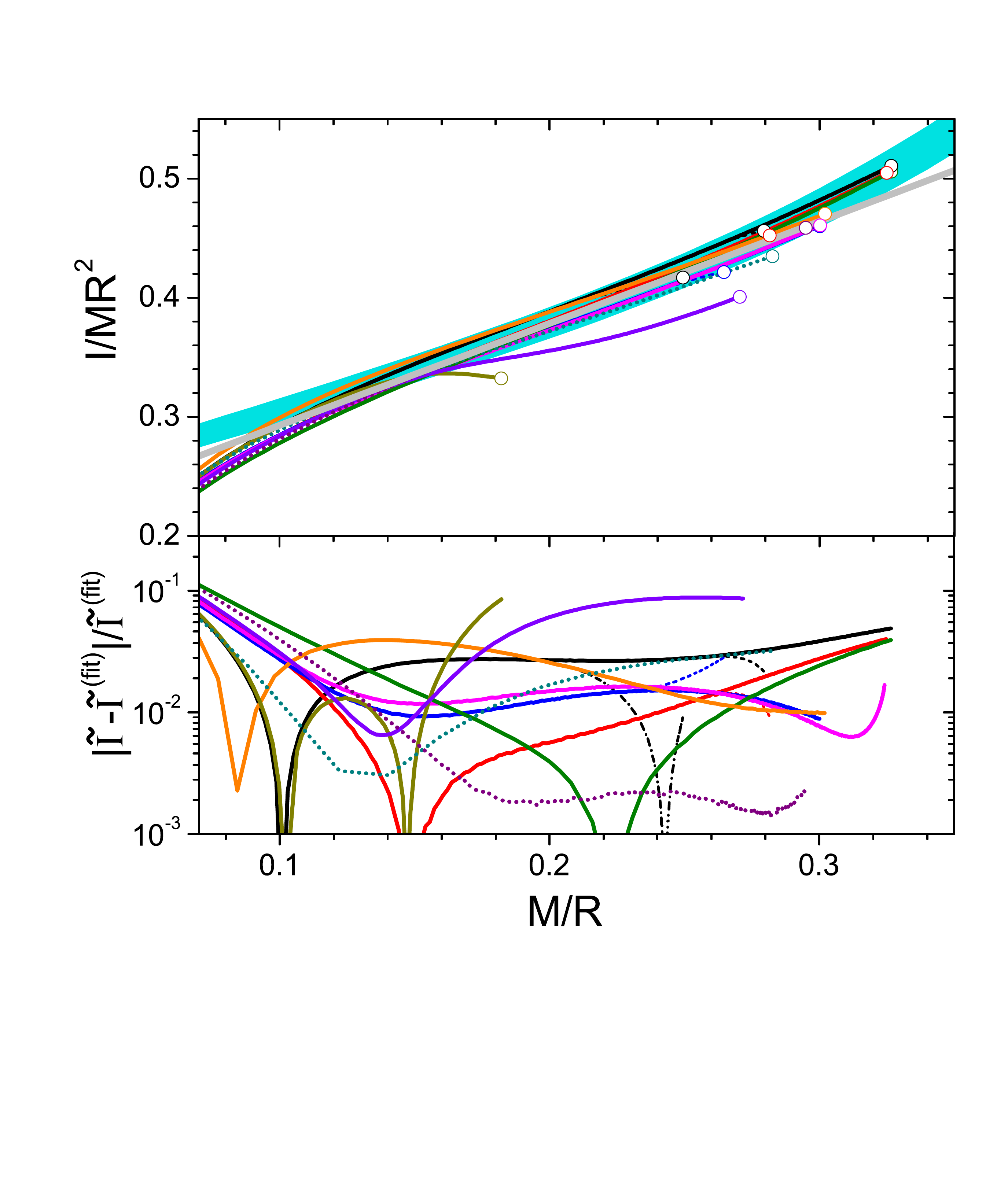}
\vspace{-27mm}
\caption{
$I/M^3$ (left panel) and $I/MR^2$ (right panel) vs.~$\be=M/R$
for 10+4 different nucleonic+hybrid EOSs.
The lower panels show the fractional deviations from the fit curves.
}
\label{f:i}
\end{figure}

In Fig.~\ref{f:i} (left panel)
we show the normalized moment of inertia $I/M^3$
vs.~the compactness $\be$
for the same EOSs as in Fig.~\ref{f:mr}.
We see that the curves obtained with different EOSs are nearly EOS-independent.
The universal relation displayed in the left panel
is the one advocated in \cite{breu},
\beq
 \frac{I}{M^3} = 0.8134\beta^{-1} + 0.2101\beta^{-2}
 + 0.003175\beta^{-3} - 0.0002717\beta^{-4}  \:.
\label{e:breu}
\eeq
As clearly shown, our chosen set of microscopic EOSs
fulfills very well the universal relation.
On the other hand, the right panels of Fig.~\ref{f:i} show the results obtained
for the quantity $I/MR^2$, together with a simple linear fit
which for our chosen set of microscopic EOSs reads
\beq
 \frac{I}{MR^2} = 0.207 + 0.857\beta \pm 0.011 \:,
\eeq
to be compared with the one reported in
Ref.~\cite{lattimer2001,steiner2010,steiner2013},
obtained with a larger set of EOSs, and displayed as a blue band,
\beq
 \frac{I}{MR^2} =
 (0.237 \pm 0.008) (1 + 2.844\beta + 18.91\beta^4) \:.
\label{e:fitmb}
\eeq
It may be noted that the fit fails for hyperonic stars with low $\mmax$.
This feature is caused by the small radius of the
maximum mass configuration for the hyperonic EOS, see Fig.~\ref{f:mr},
which leads to an `abnormaly' large value of $\beta=M/R$ close to the
(small) maximum mass.
Turning the universality argument around,
future simultaneous measurement of $M,R,I$
could therefore provide evidence for the presence of hyperons in a NS,
at least close to the maximum mass configuration.
Summarizing, in both figures
the deviations of the individual EOSs from the universal fits, shown in the
lower panels, are of the order of a few percent,
largest with the hyperonic EOS.

A further universal, i.e., EOS-independent,
relation involves the moment of inertia $I$ and the spin-induced
quadrupole moment $Q$,
as proposed in Refs.~\cite{yagi13,yagi17}.
We found that our set of microscopic EOSs fulfills also this universal relation,
as discussed in Ref.~\cite{2018arXiv180904315W}.

\section{Conclusions}
\label{s:end}

We have demonstrated that the microscopic EOSs
derived some time ago in the BHF formalism
based on meson-exchange nucleon-nucleon potentials
and consistent microscopic three-body forces,
are fully compatible with new constraints imposed by interpretation of the
first observed NS merger event GW170817.
Moreover they are consistent with transverse flow data on heavy ion reactions,
and subthreshold kaon production.
In particular, they respect the lower $2\ms$ limit of the NS maximum mass
and feature typical radii between 12 and 13 km,
constrained by tightly correlated values of the tidal deformability $\la$.
The same holds true also for the APR EOS, the relativistic DBHF,
and the two phenomenological relativistic-mean-field EOSs, LS220 and SFHo.
We have also confirmed the validity of several universal relations
among the moment of inertia $I$,
the tidal deformability $\la$,
and the quadrupole moment $Q$,
as proposed some years ago \cite{yagi13,yagi17}.

The detection of gravitational waves by the LIGO/Virgo collaboration in 2015,
and the successive binary neutron star merger event GW170817,
opened a new astronomical eye to the Universe,
and NSs play a key role in this respect,
having the potential of being extremely prolific gravitational wave emitters
in terms of expected detection rates.
Therefore we are looking forward to more refined constraints
to be obtained soon from further merger events and new facilities.

\section{Acknowledgements}

J.-B. Wei acknowledges the China Scholarship Council
(CSC File No.~201706410092) for financial support.
Partial support comes also from ``PHAROS," COST Action CA16214.

\newcommand{\apjl}{Astrophys. J. Lett.\ }
\newcommand{\physrep}{Phys. Rep.\ }
\newcommand{\mnras}{Mon. Not. R. Astron. Soc.\ }
\newcommand{\aap}{Astron. Astrophys.\ }
\newcommand{\prc}{Phys. Rev. C \ }
\newcommand{\prd}{Phys. Rev. D \ }
\newcommand{\prl}{Phys. Rev. Lett. \ }

\bibliographystyle{JHEP}
\bibliography{ConfXIII}

\end{document}